# Meta-optic Accelerators for Object Classifiers

Hanyu Zheng[1], Quan Liu[2], You Zhou[4], Ivan I. Kravchenko[5], Yuankai Huo[2], and Jason Valentine[3*].

1. Department of Electrical and Computer Engineering, Vanderbilt University, Nashville, TN, 37212
2. Department of Computer Science, Vanderbilt University, Nashville, TN, USA, 37212.
3. Department of Mechanical Engineering, Vanderbilt University, Nashville, TN, USA, 37212.
4. 1Interdisciplinary Materials Science Program, Vanderbilt University, Nashville, TN, 37212, USA
5. Center for Nanophase Materials Sciences, Oak Ridge National Laboratory, Oak Ridge, TN, USA, 37830

* Corresponding author: jason.g.valentine@vanderbilt.edu

**Abstract**

Rapid advances in deep learning have led to paradigm shifts in a number of fields, from medical image analysis to autonomous systems. These advances, however, have resulted in digital neural networks with large computational requirements, resulting in high energy consumption and limitations in real-time decision making when computation resources are limited. Here, we demonstrate a meta-optic based neural network accelerator that can off-load computationally expensive convolution operations into high-speed and low-power optics. In this architecture, metasurfaces enable both spatial multiplexing and additional information channels, such as polarization, in object classification. End-to-end design is used to co-optimize the optical and digital systems resulting in a robust classifier that achieves 95% accurate classification of handwriting digits and 94% accuracy in classifying both the digit and its polarization state. This approach could enable compact, high-speed, and low-power image and information processing systems for a wide range of applications in machine-vision and artificial intelligence.

## Introduction

Digital neural networks (NNs) and the availability of large training datasets have allowed for rapid progress in the performance of machine-based tasks for a wide range of applications including image analysis[1, 2], sound recognition[3, 4], and natural language translation[5]. The enhanced capability has, however, come at a computational cost as increased complexity and accuracy have necessitated the need for ever larger deep neural networks (DNNs)[6]. The ever-increasing computational requirements of DNNs have resulted in unsustainable growth in energy consumption and restrictions in real-time decision making when large computational systems are not available.

One alternative to DNNs is the use of optical processors that have the advantages of ultra-fast processing times and low energy costs[7, 8, 9]. These systems can be employed as stand-alone processors or as front-end accelerators for digital systems. In either case, optical systems are most impactful when used for the linear matrix-vector multiplications[10, 11] that comprise the convolution operations in DNNs as these are often the most computationally burdensome components typically comprising more than 90% of the required floating-point operations (FLOPs) in popular CNNs[12, 13]. There are both free-space[14, 15, 16]and chip-based[17, 18] approaches to optical processors but in either case, the computational advantage is achieved via the massively parallel and low power processing that is possible with optics. In the case of image analysis, free-space approaches are attractive as spatial multiplexing can be readily achieved[19, 20, 21]as well as the fact that an optical front-end can potentially be integrated directly with an imaging system[22, 23].

The most traditional approach to free-space based optical image processing is the use of $4f$ optical correlators where spatial filters[24, 25, 26, 27], either passive or dynamic, are placed in the Fourier plane of a 2-lens optical system. Recorded spatial features are then fed to a lightweight digital NN back-end for classification. An alternative approach is the use of diffractive neural networks which utilize cascaded diffractive elements as convolutional layers[28, 29, 30]. Image classification is realized through redistribution of optical energy on the detector plane requiring minimal digital processing. The tradeoff is the need for several diffractive layers as well as coherent illumination, precluding use with ambient lighting. While these approaches have shown benefits in terms of processing speed and energy consumption, they necessitate enlarged imaging systems. Furthermore, none of these approaches utilize the additional information channels, such as polarization, that are available when utilizing an optical front-end[31, 32, 33].

Here, we demonstrate the use of meta-optic based optical accelerators that serve as the convolutional front-end for a hybrid image classification system. Spatial multiplexing is achieved by using a multi-channel metalens for image duplication and a metasurface-based convolutional layer. This system has the advantage of being compact while the use of metasurfaces allows for additional information channels[34, 35, 36], in this case, polarization, to be accessed enabling both image and polarization-based classification. The hybrid network utilizes end-to-end design such that the optical and digital components are co-optimized while also incorporating statistical noise resulting in a robust classification network. We experimentally demonstrate the classification of the MNIST dataset[37] with an accuracy of 95% as well as 94% accurate classification of polarized MNIST digits. Due to the compact footprint, ease of integration with conventional imaging systems, and ability to access additional information channels, this type of system could find uses in high-dimensional imaging[38], information security[39], and machine vision[40, 41].

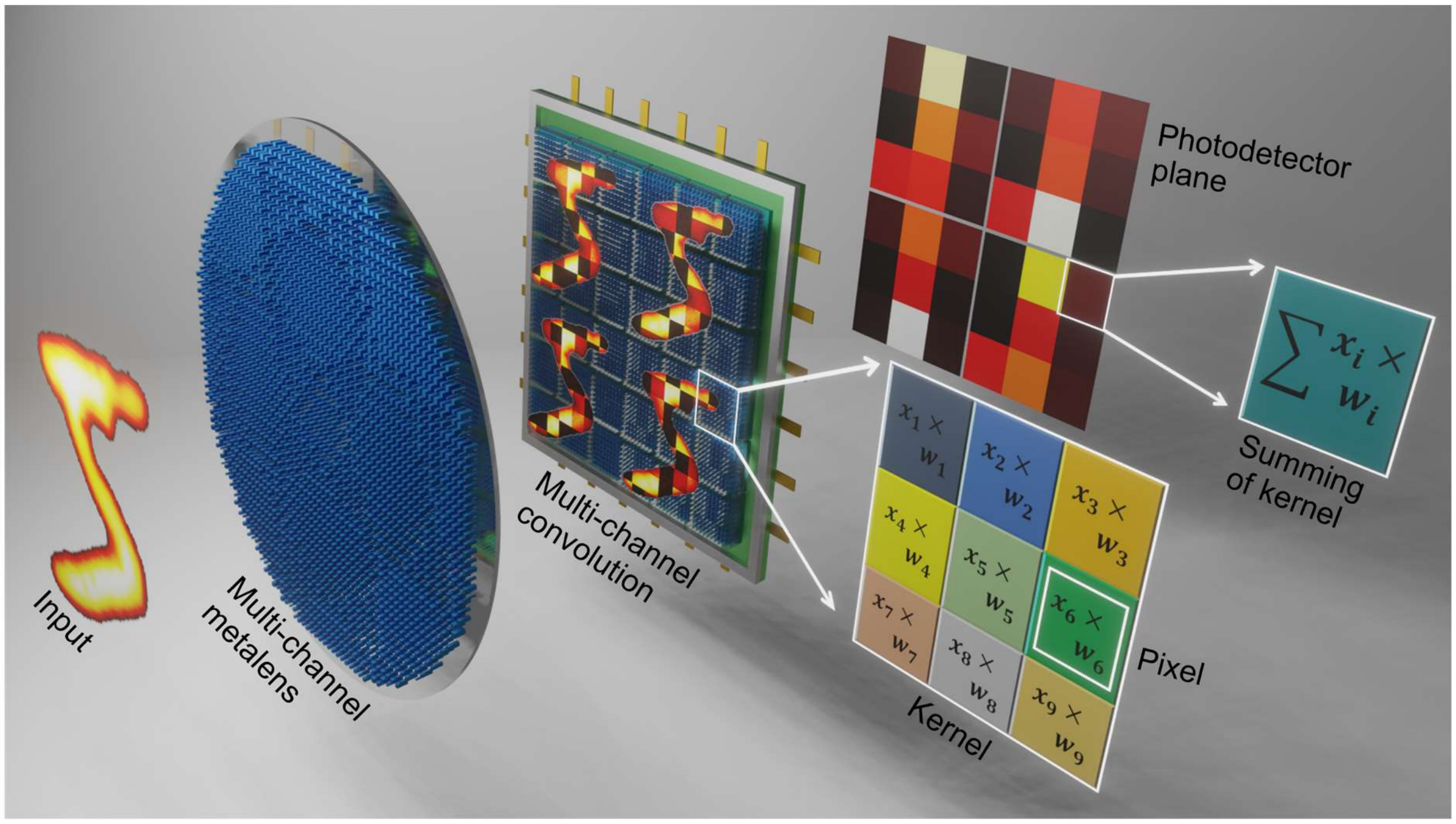

**Figure 1.** Comparison of the digital and optical convolutional neural network. The meta-optic enables multi-channel signal processing for replacing convolution operations in a digital neural network. Summing is achieved by each kernel being recorded by a single pixel on the photodetector.

## Results and Discussion

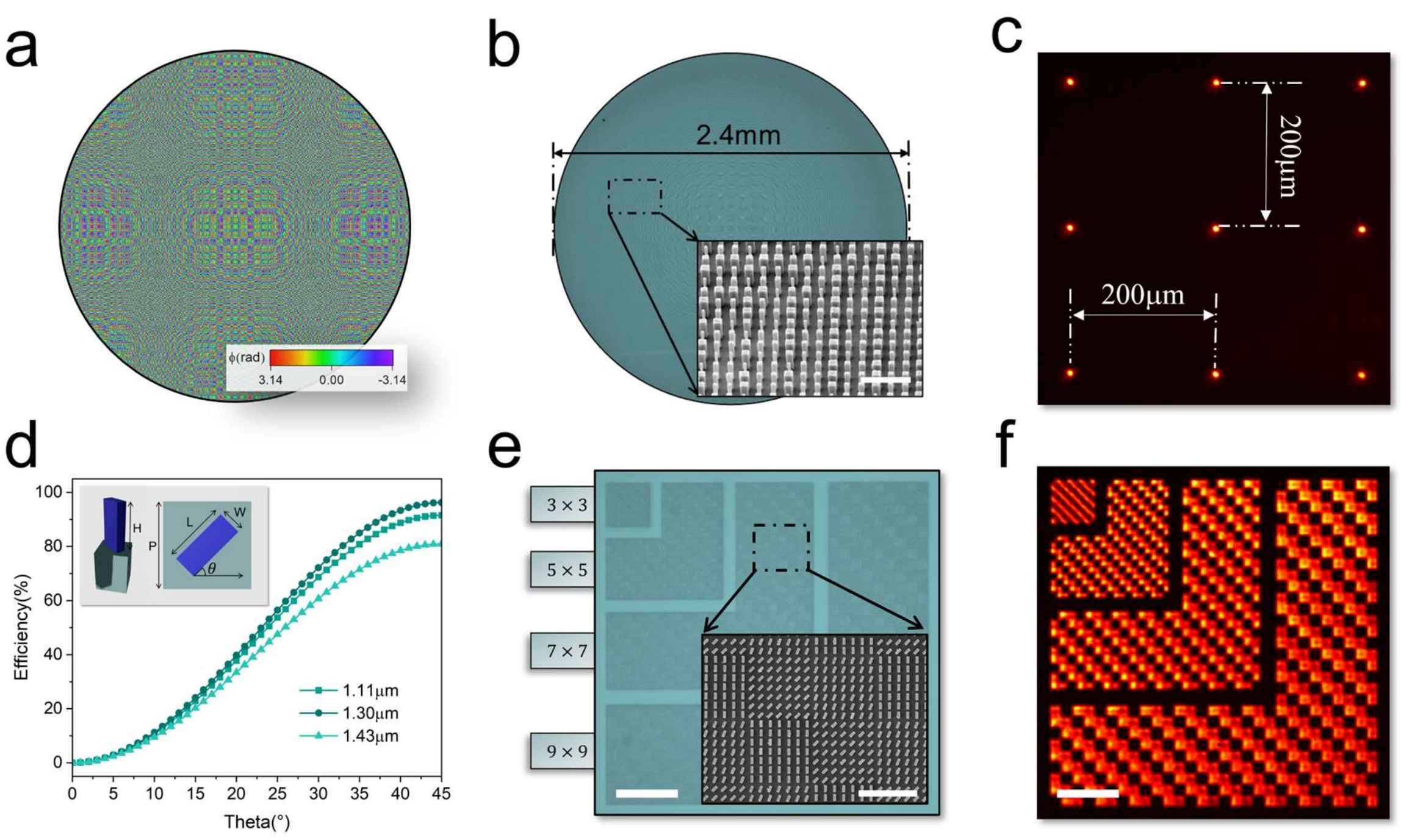

**Figure 2.** Components of the meta-optic system. (a) Phase profile of the multi-channel metalens which was achieved using spatially multiplexed meta-atoms. (b) Optical image of the fabricated multi-channel metalens. The inset is an SEM image of the lens. Scale bar: 2μm. (c) Experimental focal spot profile of the multi-channel metalens under illumination at 1.3μm. (d) Polarization conversion efficiency as a function of the meta-atom rotation angle. (e) Optical image of kernel resolution test chart. The left-side bar shows the number of meta-atoms in each kernel pixel. Inset is an SEM image of the meta-atoms. Scale bar: 4μm. (f) Optical transmission of the kernel resolution test char under 1.3μm linearly polarized laser illumination. An orthogonal analyzer was used to image the transmission map.

The meta-optic accelerator is made up of two metasurfaces, a platform chosen due to the fact that it offers precise wavefront[42], complex-amplitude[43], and polarization state[44] manipulation in an ultra-thin form factor. Metasurfaces have also been utilized as standalone systems for all-optical image processing, namely, edge detection[45], through manipulation of the non-local, angle-dependent, response. In our design, the first metasurface is a multi-channel metalens that duplicates an object into nine images as shown in Fig. 1. The multi-channel metalens was created using 9 meta-atoms per super-cell to create images at 9 spatial locations. The lens was created with a hyperbolic phase profile where the phase delay of each resonator, $i$, in the supercell is given by,

$$\phi_i = \frac{2\pi}{\lambda}\left(f - \sqrt{f^2 + (x - a_i)^2 + (y - b_i)^2}\right) \tag{1}$$

where $f$ is the focal length, $\lambda$ is the working wavelength, and $x$ and $y$ are the spatial positions on the lens. $a$ and $b$ correspond to the displacement of each unique focal spot, $i$, from the center of the lens. The resulting phase profile, for a 2.4 mm diameter metalens, is shown in Fig. 2(a). The metalens was realized using silicon nanopillars with a period of 0.6μm and a height of 0.88μm. The diameter of each meta-atom was chosen such it provides the phase profile given by Eq. 1.

Fabrication of the metalens began with a silicon device layer on quartz follow by standard electron beam lithography (EBL) followed by reactive-ion-etching (RIE). An optical image of the meta-lens is shown in Fig. 2(b) with the inset showing the individual meta-atoms. The experimentally recorded focal spots demonstrate diffraction limited performance, as shown in Fig. 2(c). While spatial multiplexing is used here to create the multi-channel lens, it is worth noting that the design method is not unique. As additional channels are added a spatially multiplexed lens will suffer from higher order diffraction and resolution reduction due to a larger super-cell structure. One way this can be overcome is through the use of complex-valued amplitude modulation which results in a smaller meta-atom supercell. In this technique, even a $2 \times 2$ meta-atom unit cell can be used for arbitrary channel metalenses, allowing the spatial resolution of each image to be maintained, as shown in the supplementary information.

The second metasurface serves as a multifunctional kernel layer that provides the vector-matrix multiplication operation. The kernels are based on Pancharatnam-Berry metasurfaces[46, 47] that can encode polarization and/or amplitude information for convolution with the image. The transmission of each nanopillar comprising the metasurface follows an analytical model based on the Jones matrix given by,

$$\begin{bmatrix} E_{x,out} \\ E_{y,out} \end{bmatrix} = \begin{bmatrix} cos(\theta) & sin(\theta) \\ -sin(\theta) & cos(\theta) \end{bmatrix} \begin{bmatrix} e^{i\phi_x} & 0 \\ 0 & e^{i\phi_y} \end{bmatrix} \begin{bmatrix} cos(\theta) & -sin(\theta) \\ sin(\theta) & cos(\theta) \end{bmatrix} \begin{bmatrix} E_{x,in} \\ E_{y,in} \end{bmatrix} \tag{2}$$

where $E_{x,in}$, $E_{y,in}$ and $E_{x,out}$, $E_{y,out}$ are the $x$ and $y$ polarized incident and transmitted amplitude, respectively. $\phi_x$ and $\phi_y$ are the phase shifts provided by the resonator for $x$ and $y$ polarization, values that are dictated by the size of the resonator. $\theta$ is the pillar rotation angle, which determines the polarization conversion efficiency for a given pixel in the metasurface. In order to control the weights in each kernel we utilize linearly polarized incident light combined with an orthogonal polarizer, serving as an analyzer, that is placed in front of the camera. The rotation angle of each meta-atom, $\theta$, dictates the percentage, or weight, of the incident light that has had its polarization vector rotated by 90°, thus passing the analyzer. In order to achieve amplitude modulation, spatial variations in $\phi_x$ and $\phi_y$ are not needed and were fixed as $|\phi_y - \phi_x| = \pi$ for each unit cell to simplify the model. In this case, the intensity of $y$-polarized transmitted light, assuming $x$-polarized incident light, is given by,

$$I_{y,out} = sin(2\theta)^2 \cdot I_{x,in} \tag{3}$$

where $I_{x,in}$, $I_{y,in}$ and $I_{x,out}$, $I_{y,out}$ are the $x$ and $y$ polarized incident and transmitted intensities, respectively. The use of pillar rotation for controlling kernel weight has the advantage of being broadband while also allowing for precise control over the weight as rotation is readily controlled in the lithography process. Fig. 2(d) displays the transmission, $T_{yx} = I_{y,out}/I_{x,in}$ as a function of rotation angle and wavelength, revealing a 320nm bandwidth where there is less than a 10% variation in transmission. In this approach, either the camera pixel size, or the kernel size, determines the maximum areal density of neurons. In the case of the kernel, the meta-atoms in each pixel of the kernel are designed as being periodic. Thus, as the number of meta-atoms in each uniform pixel are reduced there will be deviation in the weight as boundaries of the pixels, where periodicity is broken, play a larger role. In Fig. 2(e) and (f), we characterize the role of pixel size on the accuracy of the designed weight using 3 x 3 pixel kernels and finding that a minimum pixel size of 0.2 pixels / $\lambda^2$ is possible based on a maximum weight error of 10% where $\lambda$ is the working wavelength. In this case of 1.3μm illumination this yields a minimum pixel size of 3μm x 3μm (5

x 5 meta-atoms) or ~1x$10^5$ pixels/mm$^2$. This can be compared to state-of-the-art spatial light modulators (DLP650LNIR, Texas Instruments Inc.) where pixel sizes are on the order of 10.8μm x 10.8μm yielding 9x$10^3$ pixels/mm$^2$. Understanding the minimum metasurface pixel size is also important in the case of reconfigurable metasurface kernel layers as weight must be accurately controlled regardless of the kernel pattern.

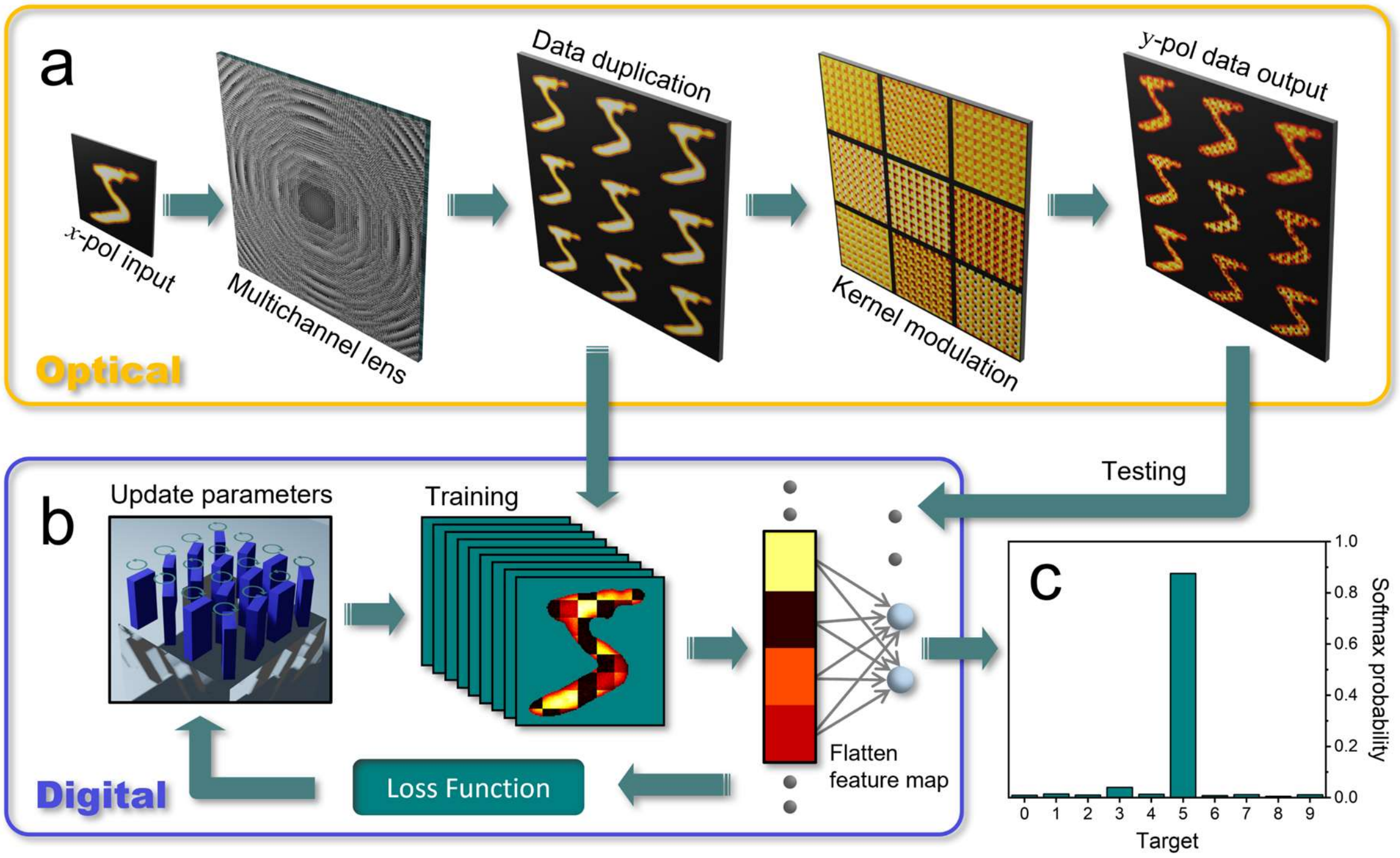

**Figure 3.** Workflow of the meta-optic accelerator design and testing process. (a) The meta-lens, with 9 independent channels, splits the image. In training, these images are recorded on the camera and transferred to the digital backend for neural network training. Once the kernel layer is trained, it is inserted into the system and the images are then projected onto that layer for testing. The transmission of the kernel layer generates 9 feature maps which are then used for testing of the network (b) The digital loop comprises an iterative training process in which the Jones matrix is used for forward propagation. (c) The probability histogram is the final output for image classification.

In order to design the weights and geometry of the kernel layer, we employ end-to-end design where both the digital and optical systems are co-optimized as shown in Fig. 3. The system was designed for classification of $24 \times 24$ pixel MNIST digits using the 9 unique channels provided by the metalens, each channel comprising $3 \times 3$ pixel kernels with a stride of 3. Implementing an optical front-end imposes unique constraints on the design of hybrid neural network structures as there are several noise sources in the analogue signal being input, and output, from the optical system. The main sources of noise in our system come from stray light, detector noise, image misalignment due to variations in the optical system, and fabrication imperfections in the metalens and kernel layers. To better understand these noise sources, and validate the designs for statistically relevant data sets, an image projection system was built comprising a spatial light modulation (SLM), illuminated with an incoherent tungsten filament lamp, for projecting the $24 \times 24$ pixel MNIST digits. The SLM was imaged, using the meta-optic, onto an InGaAs focal plane array that was triggered by the SLM such that large numbers of images could be recorded in an automated fashion.

To account for noise in the projection, imaging, and detector systems the 10,000 training images from the MNIST dataset were projected and recorded using the metalens as the imaging optic, without the kernel layer, as shown in Fig. 3(a). The optically recorded data were used as the training data in the end-to-end design loop. In the training process we incorporated 10% spatial intensity fluctuation in the kernel layer and random image rotation within ±5 degrees. The feature maps, which correspond to the convolution of the metasurface kernel layer with each of the 9 images, were fed into the trainable model to form a mean-square-error (MSE) loss function as shown in Fig. 3(b). The backward propagation comprised a stochastic gradient descent (SGD) based algorithm driven by the loss function to update the physical parameters ($\phi_x$, $\phi_y$ and $\theta$) of metasurface kernel layer for each iteration.

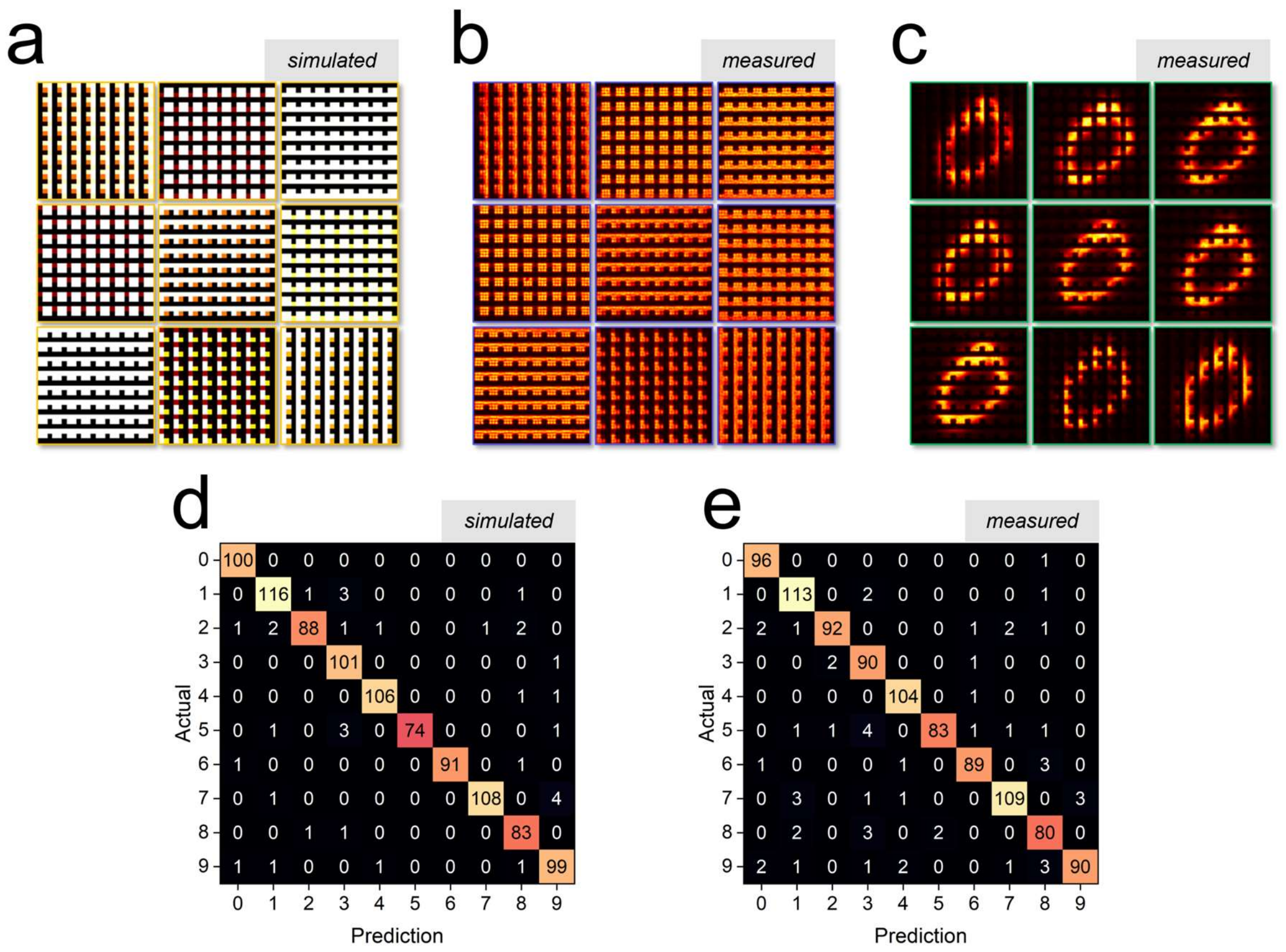

**Figure 4.** Classification of MNIST digits. (a) Transmission (weights) of the ideal kernels after the training process. (b) The measured transmission of the fabricated channels. An incoherent light source, filtered at 1.3μm, was used for illumination. (c) The 9 feature maps recorded on the camera. (d) and (e) the theoretical and measured confusion matrices for MNIST object classification.

Once training of the system was complete, the metasurface kernel layer was realized by using a silicon device layer on quartz with the device layer patterned into nanopillars with a period of 0.6μm and a height of 0.88μm. The metasurface was fabricated using EBL patterning followed by RIE. The width and length of each nanopillar was fixed as 160nm and 430nm, respectively, with the rotation angle dictated by the desired transmission value. The metasurface kernel layer was imaged using uniform illumination and compared to the theoretical design, both of which are included in Fig. 4(a) and (b). The fabricated and designed kernels show agreement with a standard deviation of less than 10%, which matches the noise level in the training model. The kernel layer was then placed in the image plane of the metalens for recording the convoluted images in the testing dataset. Fig. 4(c) shows the feature map produced for a digit of "0". Each kernel pixel comprises an $11 \times 11$ ($6.6\mu m \times 6.6\mu m$) meta-atom array. Summing of each kernel could be achieved optically via alignment of each kernel with an individual pixel on the camera, however, in this work summing is performed digitally as the kernel layer is magnified when imaged onto the camera such that each kernel comprises multiple camera pixels.

In order to characterize the system's performance, 1000 digits, not in the training set, were recorded using the meta-optic. The theoretical and experimental confusion matrices for this testing dataset are shown in Fig. 4 (d) and (e), respectively. The theoretical training model's overall accuracy was 96%, while the experimental accuracy is 95%. While this proof-of-concept demonstration involves low resolution images, the small minimum pixel size of the kernel layer along with the parallel nature of the optical operations[48, 49] means that this architecture could be a powerful tool for high-speed and large-scale image processing applications. Moreover, the versatility of the system can be further improved by incorporation of dynamically tunable metasurfaces[50] as the kernel layer such that the optical front-end can be reconfigured or temporally multiplexed.

One of the unique strengths of metasurfaces, compared to conventional lenses or diffractive optical elements, is their ability to provide user-specified amplitude and phase functions while also being sensitive to the polarization state and wavelength of light. This allows for access to additional information carriers that are normally lost when recording an image on a camera allowing one to discriminate based on normally hidden features in the physical world such as vectorial polarization, phase gradients, or spectrally complex signals. To demonstrate this ability, a polarized MNIST dataset with 8000 images was created comprising four digits (1, 4, 5, 7) with each digit having two orthogonal polarization states, as shown in Fig. 5(a). As shown in the supplementary materials, more complex, full vectorial signal recognition is also achievable using this approach. Polarization classification is possible due to the fact that the meta-atoms, outlined in Fig.2 (d), have a transmitted intensity that is dependent on the incident polarization state, as illustrated in following equation,

$$I_{y,out} = cos(2\theta)^2 \cdot I_{y,in} + sin(2\theta)^2 \cdot I_{x,in} \quad (4)$$

where $I_{x,in}$, $I_{y,in}$ and $I_{y,out}$ are the $x$ and $y$ polarized incident and transmitted intensities and $\theta$ is the meta-atom rotation angle. The optical kernel layer was design following the training procedure outlined in Fig. 3 with the system classifying eight output states comprising the four distinct digits, each with $x$ and $y$ polarization states. The metasurface layer was formed from silicon nanopillars using the same geometry and fabrication process as described previously. In Fig.5 (c) and (d) the transmitted intensity, $I_{y,out}$, is provided for a uniformly illuminated kernel layer with both $x$ and $y$-polarization states and in Fig. 5 (e) and (f) we provide the feature maps for an identical digit with $x$ and $y$-polarization states. Both the uniformly illuminated kernels and feature maps demonstrate the contrast in the convolution for orthogonal polarization states. In Fig. 5(g) and (h) we provide the theoretical and experimental confusion matrices, respectively, for 1000 test images not in the training dataset. The theoretical accuracy of classification was 95% shown while the experimental accuracy was 94%, showing excellent agreement.

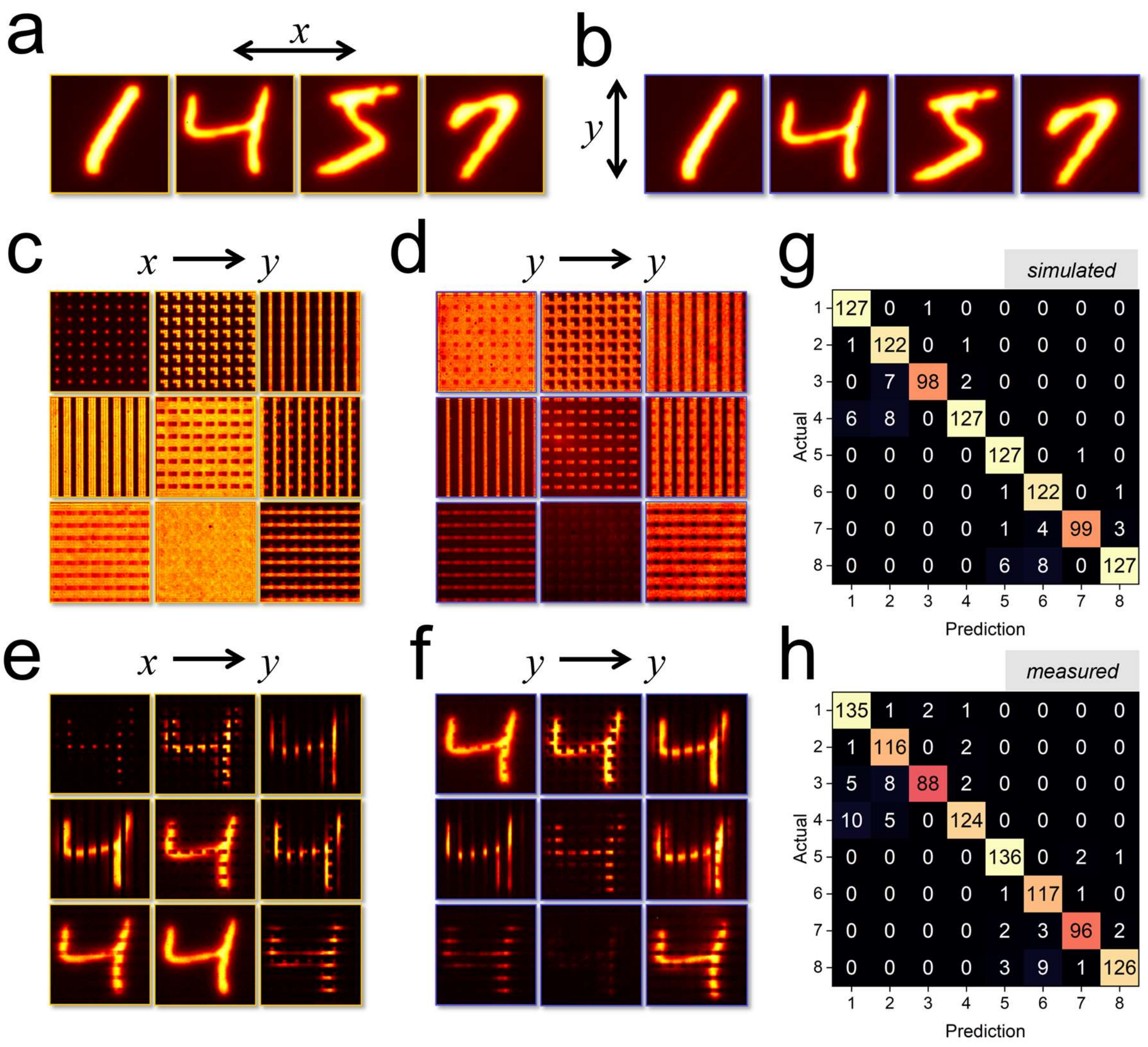

**Figure 5.** Demonstration of multi-functional object recognition. (a) and (b) illustrate the digits for classification which carry both amplitude and polarization information. (c) and (d) are the transmission profiles of the fabricated metasurface channels under the illumination of 1.3μm incoherent light with linear polarization along orthogonal directions. (e) and (f) feature maps for identical digits with *x* and *y*-polarization, respectively. (g) and (h) are the theoretical and measured confusion matrices.

## Conclusion

In summary, we have demonstrated a meta-optic accelerator for multi-functional image recognition. The technique is enabled by the unique design freedom afforded by metasurfaces, including the creation of multi-channel lenses to duplicate information and polarization sensitive kernel layers which allow for discrimination based on both the spatial intensity profile and the polarization state of the object. The use of polarization demonstrates how optical front-ends can access additional information channels normally lost in traditional imaging systems. Furthermore, by implementing end-to-end design we were able to make the system robust to common noise sources resulting in 95% accurate experimental classification that closely matches the theoretical prediction. These meta-optic accelerators demonstrate improvements in processing speed while reducing power consumption, all while opening the door to new information channels. As such, we believe they could see use in a wide range of compact, low power, and high speed computer vision systems.

## Author contributions

J.V. and H.Z developed the idea. H.Z. conducted the optical modelling and system design. Q.L and H.Z. performed the digital neural network design. Y.Z designed the multichannel metalens. H.Z fabricated the samples. I.I.K. performed the silicon growth. H.Z. performed the experimental measurements with the assistance of Y.Z. H.Z and Q.L. performed the measurement data analysis. H.Z. and J.V. wrote the manuscript with input from all the authors. The project was supervised by Y.H. and J.V.

## Acknowledgements

J.V. and H.Z. acknowledge support from DARPA under contract HR001118C0015. Part of the fabrication process was conducted at the Center for Nanophase Materials Sciences, which is a DOE Office of Science User Facility. The remainder of the fabrication process took place in the Vanderbilt Institute of Nanoscale Science and Engineering (VINSE) and we thank the staff, particularly K. Heinrich, for their support.

## Competing financial interest

The authors declare no competing financial interests.

## Reference

1. Simonyan, K. & Zisserman, A. Very deep convolutional networks for large-scale image recognition. *3rd Int. Conf. Learn. Represent. ICLR 2015 - Conf. Track Proc.* 1–14 (2015).

2. Wang, G. *et al.* Interactive Medical Image Segmentation Using Deep Learning with Image-Specific Fine Tuning. *IEEE Trans. Med. Imaging* **37**, 1562–1573 (2018).

3. Furui, S., Deng, L., Gales, M., Ney, H. & Tokuda, K. Fundamental technologies in modern speech recognition. *IEEE Signal Process. Mag.* **29**, 16–17 (2012).

4. Sak, H., Senior, A., Rao, K. & Beaufays, F. Fast and accurate recurrent neural network acoustic models for speech recognition. *Proc. Annu. Conf. Int. Speech Commun. Assoc. INTERSPEECH* **2015-January**, 1468–1472 (2015).

5. Devlin, J., Chang, M. W., Lee, K. & Toutanova, K. BERT: Pre-training of deep bidirectional transformers for language understanding. *NAACL HLT 2019 - 2019 Conf. North Am. Chapter Assoc. Comput. Linguist. Hum. Lang. Technol. - Proc. Conf.* **1**, 4171–4186 (2019).

6. Lecun, Y., Bengio, Y. & Hinton, G. Deep learning. *Nature* **521**, 436–444 (2015).

7. Wetzstein, G. *et al.* Inference in artificial intelligence with deep optics and photonics. *Nature* **588**, 39–47 (2020).

8. Shastri, B. J. *et al.* Photonics for artificial intelligence and neuromorphic computing. *Nat. Photonics* **15**, 102–114 (2021).

9. Wang, T. *et al.* An optical neural network using less than 1 photon per multiplication. 1–8 (2021) doi:10.1038/s41467-021-27774-8.

10. Xu, X. *et al.* 11 TOPS photonic convolutional accelerator for optical neural networks. *Nature* **589**, 44–51 (2021).

11. Feldmann, J. *et al.* Parallel convolutional processing using an integrated photonic tensor core. *Nature* **589**, 52–58 (2021).

12. Chen, Y. H., Krishna, T., Emer, J. S. & Sze, V. Eyeriss: An Energy-Efficient Reconfigurable Accelerator for Deep Convolutional Neural Networks. *IEEE J. Solid-State Circuits* **52**, 127–138 (2017).

13. Neshatpour, K., Homayoun, H. & Sasan, A. ICNN: The iterative convolutional neural network. *ACM Trans. Embed. Comput. Syst.* **18**, (2019).

14. Hamerly, R., Bernstein, L., Sludds, A., Soljačić, M. & Englund, D. Large-Scale Optical Neural Networks Based on Photoelectric Multiplication. *Phys. Rev. X* **9**, 1–12 (2019).

15. Mennel, L. *et al.* Ultrafast machine vision with 2D material neural network image sensors. *Nature* **579**, 62–66 (2020).

16. del Hougne, P., Imani, M. F., Diebold, A. V., Horstmeyer, R. & Smith, D. R. Learned Integrated Sensing Pipeline: Reconfigurable Metasurface Transceivers as Trainable Physical Layer in an Artificial Neural Network. *Adv. Sci.* **7**, 1–8 (2020).

17. Wu, C. *et al.* Programmable phase-change metasurfaces on waveguides for multimode photonic convolutional neural network. *Nat. Commun.* **12**, 1–8 (2021).

18. Zhang, H. *et al.* An optical neural chip for implementing complex-valued neural network. *Nat. Commun.* **12**, 1–11 (2021).

19. Zhou, Y. *et al.* Multifunctional metaoptics based on bilayer metasurfaces. *Light Sci. Appl.* **8**, (2019).

20. Wang, X., Diáz-Rubio, A. & Tretyakov, S. A. Independent Control of Multiple Channels in Metasurface Devices. *Phys. Rev. Appl.* **14**, 1 (2020).

21. Lin, Z. *et al.* End-to-end metasurface inverse design for single-shot multi-channel imaging. 1–17 (2021).

22. Li, L. *et al.* Monolithic Full-Stokes Near-Infrared Polarimetry with Chiral Plasmonic Metasurface Integrated Graphene-Silicon Photodetector. *ACS Nano* **14**, 16634–16642 (2020).

23. Reshef, O. *et al.* An optic to replace space and its application towards ultra-thin imaging systems. *Nat. Commun.* **12**, 8–15 (2021).

24. Chang, J., Sitzmann, V., Dun, X., Heidrich, W. & Wetzstein, G. Hybrid optical-electronic convolutional neural networks with optimized diffractive optics for image classification. *Sci. Rep.* **8**, 1–10 (2018).

25. Yan, T. *et al.* Fourier-space Diffractive Deep Neural Network. *Phys. Rev. Lett.* **123**, 23901 (2019).

26. Colburn, S., Chu, Y., Shilzerman, E. & Majumdar, A. Optical frontend for a convolutional neural network. *Appl. Opt.* **58**, 3179 (2019).

27. Zhou, T. *et al.* Large-scale neuromorphic optoelectronic computing with a reconfigurable diffractive processing unit. *Nat. Photonics* **15**, 367–373 (2021).

28. Lin, X. *et al.* All-optical machine learning using diffractive deep neural networks. *Science (80-. ).* **361**, 1004–1008 (2018).

29. Qian, C. *et al.* Performing optical logic operations by a diffractive neural network. *Light Sci. Appl.* **9**, (2020).

30. Luo, X. *et al.* Metasurface-Enabled On-Chip Multiplexed Diffractive Neural Networks in the Visible.

31. Khorasaninejad, M. *et al.* Multispectral chiral imaging with a metalens. *Nano Lett.* **16**, 4595–4600 (2016).

32. Arbabi, E., Kamali, S. M., Arbabi, A. & Faraon, A. Full-Stokes Imaging Polarimetry Using Dielectric Metasurfaces. *ACS Photonics* **5**, 3132–3140 (2018).

33. Rubin, N. A. *et al.* Matrix Fourier optics enables a compact full-Stokes polarization camera. *Science (80-. ).* **364**, (2019).

34. Zhao, R. *et al.* Multichannel vectorial holographic display and encryption. *Light Sci. Appl.*

**7**, (2018).

35. Kwon, H., Arbabi, E., Kamali, S. M., Faraji-Dana, M. S. & Faraon, A. Single-shot quantitative phase gradient microscopy using a system of multifunctional metasurfaces. *Nat. Photonics* **14**, 109–114 (2020).

36. McClung, A., Samudrala, S., Torfeh, M., Mansouree, M. & Arbabi, A. Snapshot spectral imaging with parallel metasystems. *Sci. Adv.* **6**, 1–9 (2020).

37. LeCun, Y., Bottou, L., Bengio, Y. & Haffner, P. Gradient-based learning applied to document recognition. *Proc. IEEE* **86**, 2278–2323 (1998).

38. Ding, F., Tang, S. & Bozhevolnyi, S. I. Recent Advances in Polarization-Encoded Optical Metasurfaces. *Adv. Photonics Res.* **2**, 2000173 (2021).

39. Kim, I. *et al.* Pixelated bifunctional metasurface-driven dynamic vectorial holographic color prints for photonic security platform. *Nat. Commun.* **12**, 1–9 (2021).

40. Li, L. *et al.* Intelligent metasurface imager and recognizer. *Light Sci. Appl.* **8**, (2019).

41. Li, L. *et al.* Machine-learning reprogrammable metasurface imager. *Nat. Commun.* **10**, (2019).

42. Kamali, S. M., Arbabi, E., Arbabi, A. & Faraon, A. A review of dielectric optical metasurfaces for wavefront control. *Nanophotonics* **7**, 1041–1068 (2018).

43. Ren, H. *et al.* Complex-amplitude metasurface-based orbital angular momentum holography in momentum space. *Nat. Nanotechnol.* **15**, 948–955 (2020).

44. Shi, Z. *et al.* Continuous angle-tunable birefringence with freeform metasurfaces for arbitrary polarization conversion. *Sci. Adv.* **6**, 1–8 (2020).

45. Zhou, Y., Zheng, H., Kravchenko, I. I. & Valentine, J. Flat optics for image differentiation. *Nat. Photonics* **14**, 316–323 (2020).

46. Khorasaninejad, M. *et al.* Metalenses at visible wavelengths: Diffraction-limited focusing and subwavelength resolution imaging. *Science (80-. ).* **352**, 1190–1194 (2016).

47. Overvig, A. C. *et al.* Dielectric metasurfaces for complete and independent control of the optical amplitude and phase. *Light Sci. Appl.* **8**, (2019).

48. Li, N. *et al.* Large-area metasurface on CMOS-compatible fabrication platform: Driving flat optics from lab to fab. *Nanophotonics* **9**, 3071–3087 (2020).

49. Zheng, H. *et al.* Large-Scale Metasurfaces Based on Grayscale Nanosphere Lithography. *ACS Photonics* **8**, 1824–1831 (2021).

50. Howes, A., Wang, W., Kravchenko, I. & Valentine, J. Dynamic transmission control based on all-dielectric huygens metasurfaces. *Opt. InfoBase Conf. Pap.* **Part F114-FIO 2018**, 787–792 (2018).

51. Hugonin, A. J. P. & Lalanne, P. RETICOLO CODE 1D for the diffraction by stacks of lamellar 1D gratings. 1–16 (2012).

52. Hughes, T. W., Minkov, M., Liu, V., Yu, Z. & Fan, S. A perspective on the pathway toward full wave simulation of large area metalenses. *Appl. Phys. Lett.* **119**, (2021).